\documentclass[12pt]{article}
\usepackage{epsfig}
\usepackage{color}

\usepackage{braket}

\newcommand{\be}{\begin{equation}}
\newcommand{\ee}{\end{equation}}

\newcommand{\bea}{\begin{eqnarray}}
\newcommand{\eea}{\end{eqnarray}}
\newcommand{\beq}{\begin{equation}}
\newcommand{\eeq}{\end{equation}}
\newcommand{\nn}{\nonumber}

\def\fun#1#2{\lower3.6pt\vbox{\baselineskip0pt\lineskip.9pt
\ialign{$\mathsurround=0pt#1\hfil##\hfil$\crcr#2\crcr\sim\crcr}}}

\begin{document}

\title{  Exotic mesons
with hidden charm as diquark-antidiquark states
}
\author{
V.V. Anisovich$^+$, M.A. Matveev$^+$, A.V. Sarantsev$^{+ \diamondsuit}$,
A.N. Semenova$^+$
}

\date{}
\maketitle

\begin{center}
{\it
$^+$National Research Centre "Kurchatov Institute",
Petersburg Nuclear Physics Institute, Gatchina, 188300, Russia}

{\it $^\diamondsuit$
Helmholtz-Institut f\"ur Strahlen- und Kernphysik,
Universit\"at Bonn, Germany}

%

\end{center}

\begin{abstract}
Exotic mesons with hidden strange ($s\bar s$) and heavy quark pairs
($Q\bar Q$), where $Q=c,b$, are considered as
diquark-antidiquark systems,
$(Qs)(\bar Q\bar s)$. Taking into account that these states
can recombinate into two-meson ones,
we study the interplay of these states in terms
of the dispersion relation D-function technique. The classification
of exotic mesons is discussed, coefficients for decay modes are given, predictions
for new states are
presented.
The nonet structure for $\Big((Qq)(\bar Q \bar q)\Big)$, $\Big((Qs)(\bar Q\bar s$)\Big),
$\Big((Qq)(\bar Q\bar s)\Big)$- states ($q=u,d$) is suggested.

\end{abstract}


\section{Introduction}

\begin{figure}
\centerline{\epsfig{file=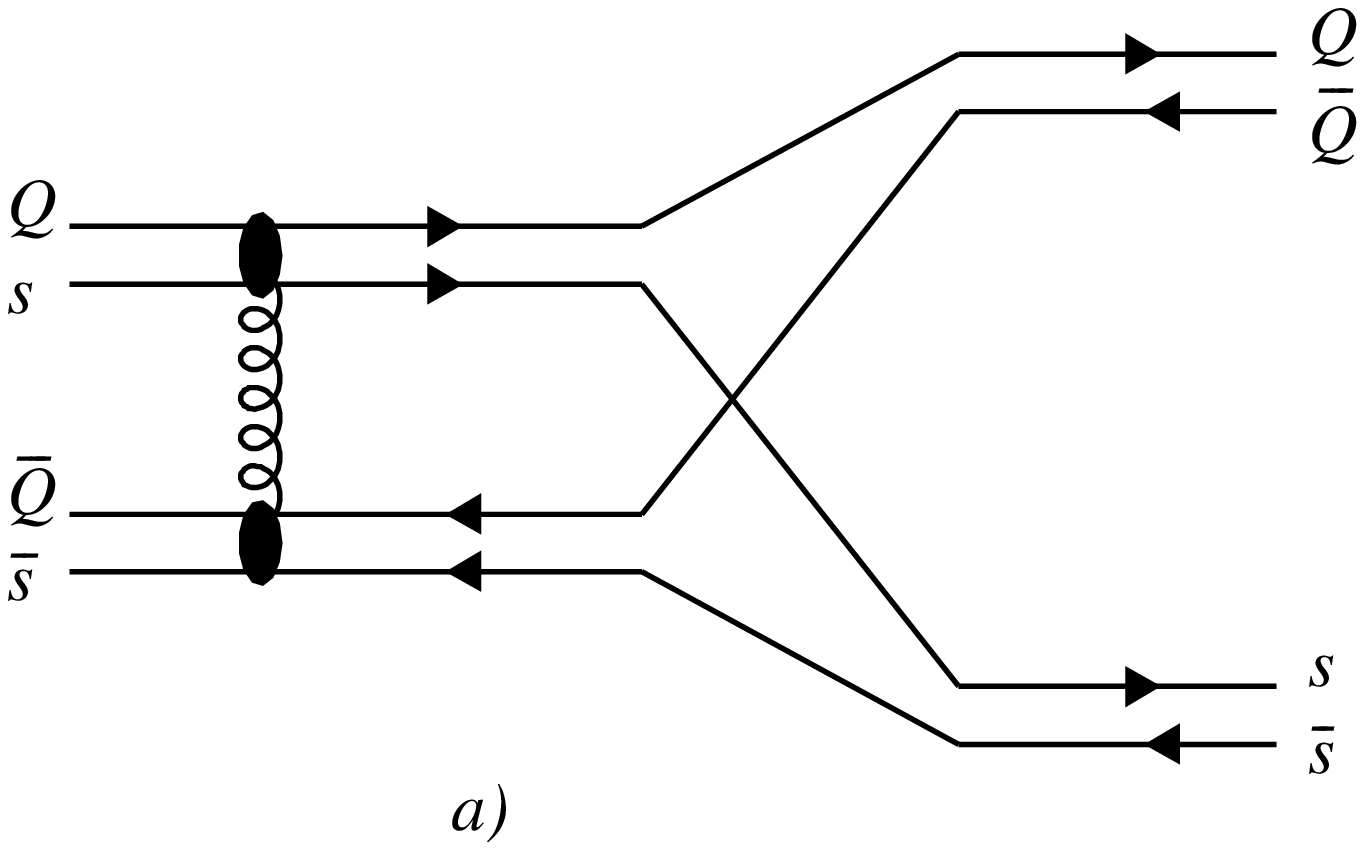,height=0.23\textwidth}\hspace{5mm}
            \epsfig{file=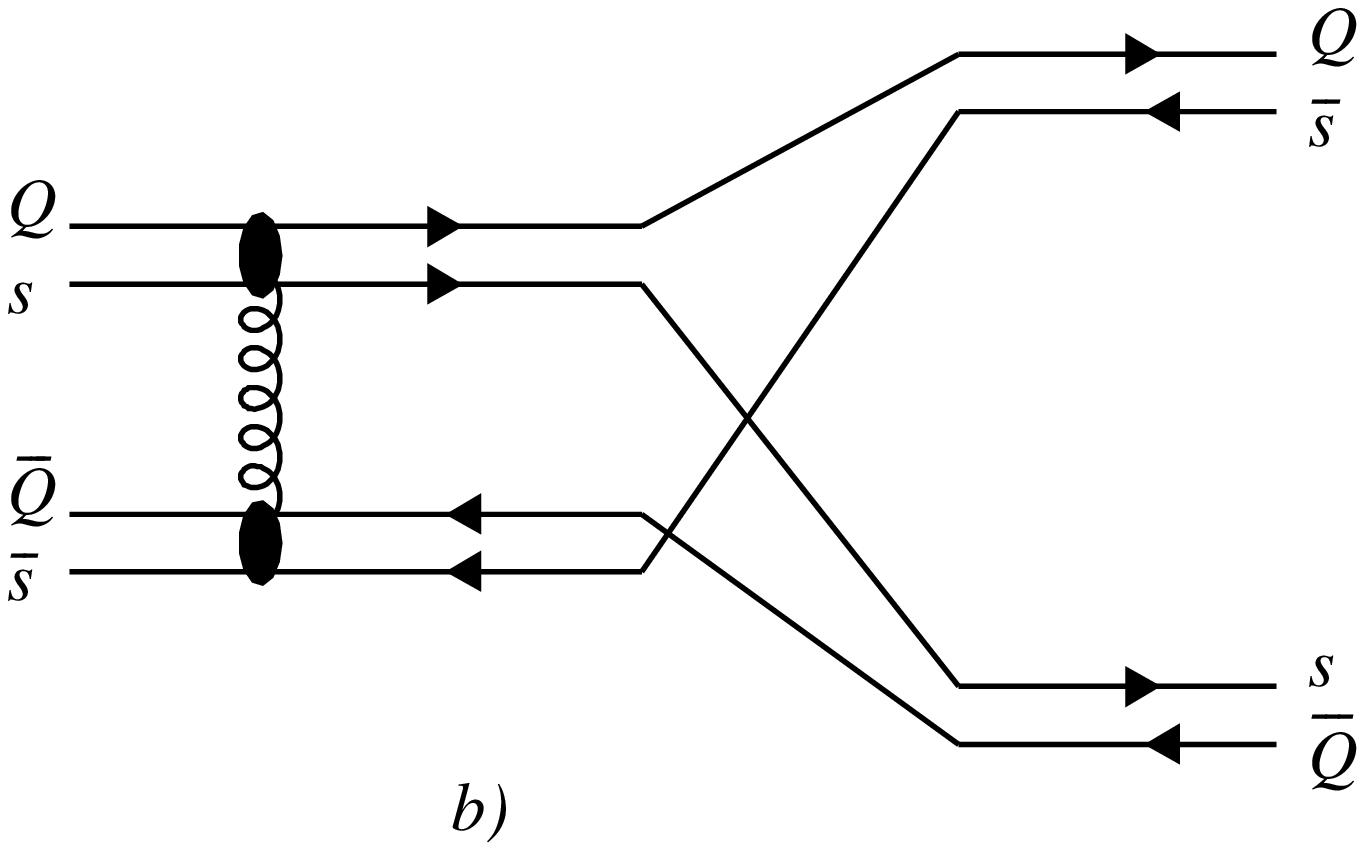,height=0.23\textwidth}}
\caption{Exotic mesons as
diquark-antidiquark states and their recombination into
mesons.
\label{f1}}
\end{figure}

Presently we have several candidates for exotic mesons in the region of masses of
charmonia $c\bar c$ and bottomonia $b\bar b$, see the
PDG-compilation \cite{PDG}. Some of them are debatable but on the
whole the observations give a strong argument to the existence of  such states.
Here we discuss a scheme in which the exotic meson states are formed
by standard QCD-motivated interactions (gluonic exchanges, confinement forces)
but with diquarks as constituents.
The formed diquark-antidiquark composite states should reveal themselves in decays
caused by the recombination processes,
see Fig. {\ref{f1}}.

The construction of standard meson states with QCD-motivated interactions is a subject
 of multitudinous studies,
without pretending for completeness we point out the refs. \cite{isgur,petry,book3}
and for the discussion of status of the meson states
refs. \cite{bugg,klempt,ufn} with
references therein.
The notion of the diquark was introduced by Gell-Mann \cite{gell-mann};
for baryon states it is discussed for
a long time, see pioneering papers \cite{ida,licht,ono,vva75,schm} and conference
presentations \cite{diquark1,goeke,diquark2}. The systematization of baryons in terms
of the quark-diquark states is presented in \cite{qD,book4}.
The extension of the diquark notion to meson states
leads to consideration of exotics, such extension in the sector of
heavy  diquarks is under discussion, see Maiani $et\,al.$ \cite{maiani}, Voloshin
\cite{voloshin}, Ali $et\, al.$ \cite{ali}.

In the paper we consider comparatively simple diquark-antidiquark systems, the
exotic meson as a product of heavy-strange color
 diquarks:
\be \label{d1}
 (Qs)\cdot(\bar Q\bar s)=Q_\alpha s_\beta\,
\varepsilon_{\alpha \beta\gamma}\,
\varepsilon_{\alpha' \beta' \gamma}\,\bar Q_{\alpha'}\bar
 s_{\beta'},
 \ee
where indices
$\alpha, \alpha', \beta, \beta', \gamma$ refer to color.
Diquarks are supposed to be effective
composite particles with constituent quarks in the S-wave that lead to
two types of diquarks, scalar and axial-vector ones:
\bea \label{d2}
{\rm scalar},\;J^P\ =\ 0^+&:&(Qs)_{0^+}\equiv S_{(Qs)}\, , \\
{\rm axial-vector},\;J^P\ =\ 1^+&:&(Qs)_{1^+}\equiv A_{(Qs)} \, .
\nn
\eea
Diquark-antidiquark states can recombinate into two mesons:
\bea \label{d3-1}
 &&
\varepsilon_{\alpha \beta\gamma}
\varepsilon_{\alpha' \beta' \gamma}=
\delta_{\alpha \alpha'}\delta_{\beta \beta'}-
\delta_{\alpha\beta' }\delta_{\beta \alpha'}\,,  \\
 &&
(Qs)\cdot(\bar Q\bar s)=(Q\bar Q)(s\bar s)-(Q\bar s)(s\bar Q).
\nn
 \eea
Therefore, the exotic states are two-component systems with diquark-antidiquark and two-meson components. The dominance of the meson-meson component means that we deal with a
molecule-like or deutron-like system  \cite{jaffe,wein,beveren}.

In this paper we consider tetraquark systems with hidden charm and hidden strangeness
and study the interplay of the diquark-antidiquark and two-meson components in terms of dispersion relation D-function technique. A classification of these meson states is suggested and coefficients for decay modes are given. Comparison with existing data
is discussed and predictions for new states are presented.
The paper is organized as follows.
In Section 2 we present a qualitative classification of the
$(Qs)\cdot(\bar Q\bar s)$-systems
following results of previous considerations \cite{maiani,voloshin,ali} and a
model treating $(Q\bar Q)$-systems \cite{ADMNS-cc,ADMNS-bb}.
Spin and orbital momentum splittings of levels are given, the effect of strange quark weighting is estimated. The effect of recombination of diquark-antidiquark states into
two-meson ones with subsequent meson-meson rescatterings play an important role in the formation of the singularities in the energy plane. We discuss the resulting singular structure
of production amplitudes in Section 3 in terms of D-function technique for meson-meson channels.
Comparison of calculations with data in the charmonium sector
(Belle \cite{belle-ss}, CDF \cite{cdf-ss}, CMS \cite{CMS-ss}, D0 \cite{D0-ss} )
is given in Section 4. The determination of the mesons from the
$(cs)\cdot(\bar{c}{\bar s})$-sector allows to fix exotic mesons in non-strange sectors,
$(cq)\cdot(\bar{c}{\bar q})$ where $q=u,d$, thus setting nonet classification for exotic mesons, Section 5.

The appendices are devoted to technicalities.
In Appendix A we present the D-function method in terms of the dispersion relation technique for final state meson-meson rescatterings.
Wave function decompositions for transitions of the diquark-antidiquark states into the
meson-meson ones is given in Appendix B.

\section{Classification of the diquark-antidiquark states   }

Here we present a qualitative classification of the
$(Qs)\cdot (\bar Q\bar s)$-systems, it is given in line with
studies of \cite{maiani,voloshin,ali} and
model calculations of the $Q\bar Q$-systems \cite{ADMNS-cc,ADMNS-bb}.  We accept that
diquarks and quarks, possessing similar color structure, allow similar model treating.
Correspondingly, we guess that
color forces result in similar mass splittings.

\subsection{Low-lying states and diquark masses }
Here we estimate masses of low-lying states keeping for a pattern
results for $(Q\bar Q)$-systems which were obtained in \cite{ADMNS-cc,ADMNS-bb}.

\subsubsection{\boldmath S-wave states}

The lowest states are S-wave composite systems of diquark and
antidiquark, they are as follows:
\be \label{4}
\begin{tabular}{l|l}
 $J^{PC}$ & $\Psi$
  \\
\hline
  $ 2^{++}$ &
$\Psi^{(AA)}_{ij}\ =\
A_{i}^{(Qs)}\cdot A_{j}^{(\bar Q\bar s)}- \frac13
\delta_{ij}\,(A_{\ell}^{(Qs)}\cdot A_{\ell}^{(\bar Q\bar s)}$)
 \\
  $ 1^{+-}$ &
$\Psi^{(AA)}_{\ell}\ =\
\frac{1}{\sqrt 2}\,\epsilon_{\ell ij}\,(A_i^{(Qs)}\cdot A_j^{(\bar Q\bar s)})$
 \\
  $ 0^{++}$ &
  $\Psi^{(AA)}\ =\
\frac{1}{\sqrt 3}(A_i^{(Qs)}\cdot A_i^{(\bar Q\bar s)})$
 \\
 $1^{++}$ &
 $\Psi^{[AS]}_{i}\ =\
\frac{1}{\sqrt 2}\Big(A_i^{(Qs)}\cdot S^{(\bar Q\bar s)}+S^{(Qs)}\cdot A_i^{(\bar Q\bar s)}$
 \Big)
 \\
 $1^{+-}$ &
 $\Psi^{\{AS\}}_{i}\ =\
\frac{1}{\sqrt 2}\Big(A_i^{(Qs)}\cdot S^{(\bar Q\bar s)}-S^{(Qs)}\cdot A_i^{(\bar Q\bar s)}$ \Big)
 \\
  $0^{++}$ &
  $\Psi^{(SS)}\ =\
(S^{(Qs)}\cdot S^{(\bar Q\bar s)})$
  \\
\end{tabular}
\ee
where the indices ($i,j,\ell$) refer to spin projections of the axial diquark.

Follow of the quark model estimations it is reasonable
to consider diquark masses of the order of:
\bea  \label{5}
&&
m_{(cq)}\sim (1550-1850)\; {\rm MeV},\qquad
 m_{(cs)}\sim (1650-1950)\; {\rm MeV},
 \\
 &&
m_{(bq)}\sim (5050-5350)\; {\rm MeV},\qquad
 m_{(bs)}\sim (5150-5450)\; {\rm MeV}.
\nn
\eea
For comparison recall the masses of charmed and beauty quarks:
$m_c\simeq 1275$ MeV \cite{PDG},
$m_c\simeq 1250$ MeV \cite{ADMNS-cc}
 and
$m_b\simeq 4650$ MeV \cite{PDG},
$m_b\simeq 4500$ MeV \cite{ADMNS-bb},
$4000<m_b(QCD)< 4500$ MeV \cite{mano}.

\subsubsection{Mass splitting of states with different spins}
Further we accept a simple mass formula which takes into account masses of constituents
and spin splitting only:
\be   \label{7}
m^{J}_{(q\bar q)}=m_{q}+m_{\bar q} +J(J+1)\Delta .
\ee
Models for $(q\bar q)$-states tell us that $\Delta=(50-100)\,{\rm MeV}$.

\subsection{Charmonium sector with hidden strangeness}

We concentrate attention to states with hidden charm and strangeness, $(c\bar c\, s\bar s)$,
- these systems can be discussed
on the basis of the
observations of collaborations Belle \cite{belle-ss}, CDF \cite{cdf-ss},
CMS \cite{CMS-ss}, D0 \cite{D0-ss} and LHCb \cite{LHCb-ss}.

\subsubsection{Exotic states with hidden charm and strangeness, \boldmath$(c\bar c\, s\bar s)$}

For charmed mesons
($D$, $D^*$, $D_s$, $D^*_s$)
the weighting of the
strange quark reads:
\be  \label{8}
m_{D_s}-m_{D}\simeq m_{D^*_s}- m_{D^*}\simeq 100\,{\rm MeV}.
\ee
We suppose the same value for the diquarks:
\be     \label{9}
m_{(cs)}-m_{(cq)}\simeq  100\,{\rm MeV}.
\ee

Further we restrict ourself by the consideration of the S-wave states.
We do not include the state $X^{(1^{--})}(4660)$ into exotics because
in this mass region there are two $(c\bar c)$ states \cite{ADMNS-cc},
namely, $\psi_{(5S)}(4570)$ and $\psi_{(4D)}(4710)$.

\subsubsection{Suggested classification of the \boldmath$cs\cdot \bar c\bar s$-states
\label{s222}}

The following states with hidden strangeness are considered as diquark-antidiquark systems:
\be \label{11}
\begin{tabular}{l|l|l|l}
& observed peak  & reaction, $\Gamma$, $J^{PC}$  &  classification
\\
\hline
 &
 &
 &  $2^{++} \quad (A_{(Qs)}\cdot A_{(\bar Q\bar s)})$
 \\
 &
 &
 &  \qquad$\sim$(4580 - 4680) MeV
\\
\hline \hline
 &
 &
 &  $1^{+-}\quad(A_{(Qs)}\cdot A_{(\bar Q\bar s)})$
 \\
 &
 &
 &  \qquad$\sim$(4380 - 4420) MeV
\\ \hline \hline
 &
 & $e^+e^-\to e^+e^- (\phi J/\psi)$,
 & $0^{++}\quad(A_{(Qs)}\cdot A_{(\bar Q\bar s)})$
\\ \cline{1-3}
   Belle \cite{belle-ss}:
 & $X_{(s\bar s)}(4350\pm 5)$,
 & $\Gamma\simeq$(5-30) MeV, $0^{++}/2^{++}$
 & \qquad$\sim$4280 MeV
\\ \hline \hline
 &
 & $ B\to K(\phi J/\psi)$,
 & $1^{++}\quad\frac{1}{\sqrt2}\Big(A_s\cdot\bar  S_s+S_s\cdot\bar A_s\Big)$
\\ \cline{1-3}
   CDF \cite{cdf-ss}:
 & $X_{(s\bar s)}(4274\pm 7)$,
 & $\Gamma\simeq$(20-50) MeV, $?^{?+}$
 & \qquad$\sim$(4310 - 4350) MeV
 \\ \cline{1-3}
   D0 \cite{D0-ss}
 & $X_{(s\bar s)}(4329\pm 12)$
 & $\Gamma\simeq$(12-62) MeV, $?^{?+}$
 &
\\ \cline{1-3}
   CMS \cite{CMS-ss}
 & $X_{(s\bar s)}(4314\pm 13)$
 & $\Gamma\simeq$(8-88) MeV, $?^{?+}$
 &
\\ \hline \hline
 &
 &
 & $1^{+-}\quad\frac{1}{\sqrt2}\Big(A_s\cdot\bar  S_s-S_s\cdot\bar A_s\Big)$
\\
 &
 &
 & \qquad$\sim$(4310 - 4350) MeV
\\ \hline \hline
 &
 & $ B\to K(\phi J/\psi)$,
 & $0^{++}\quad( S_s\cdot\bar  S_s)$
\\ \cline{1-3}
  CDF \cite{cdf-ss}:
 & $X_{(s\bar s)}(4143\pm 3)$,
 & $\Gamma\simeq$(8 - 26) MeV, $?^{?+}$
 & \qquad$\sim$4140 MeV
\\  \cline{1-3}
   D0 \cite{D0-ss}
 & $X_{(s\bar s)}(4159\pm 11)$
 & $\Gamma\simeq$(1-14) MeV, $?^{?+}$
 &
\\ \cline{1-3}
   CMS \cite{CMS-ss}
 & $X_{(s\bar s)}(4148\pm 9)$
 & $\Gamma\simeq$(28$\pm$ 30) MeV, $?^{?+}$
 &
\\
\end{tabular}
 \ee
Applying to the data \cite{belle-ss,cdf-ss,CMS-ss,D0-ss} the mass formula (\ref{7})
we write mass parameters: $m_S=2070$~MeV, $m_A=2140$ MeV,  $\Delta=(50-100)$ MeV.
Value of the mass weighting for strange quark, see eq. (\ref{9}), show to
$X(3920)$ \cite{PDG} as an analog of $X_{(s\bar s)}(4148\pm 9)$
 in the $\omega J/\Psi$ system.

The $X_{(s\bar s)}-$states have a normal width, of the order $\Gamma\sim 30$ MeV.

\section{Decomposition of the diquark-antidiquark states and
meson-meson  D-functions \label{sec3}}

It is reasonable to suggest that the recombination of quarks is responsible for the dominant
mode of the decay, see Fig. \ref{f1}.
Using the dispersion relation D-function
technique (see Appendix A and ref. \cite{book3} for more detail)
 we write resonance production amplitudes taking into account final state
meson-meson rescatterings. Correspondingly, we present here loop diagrams for meson-meson
transitions; necessary transformation of the diquark-antidiquark wave functions into
meson-meson ones is given in Appendix B.

\begin{figure}
\centerline
{\epsfig{file=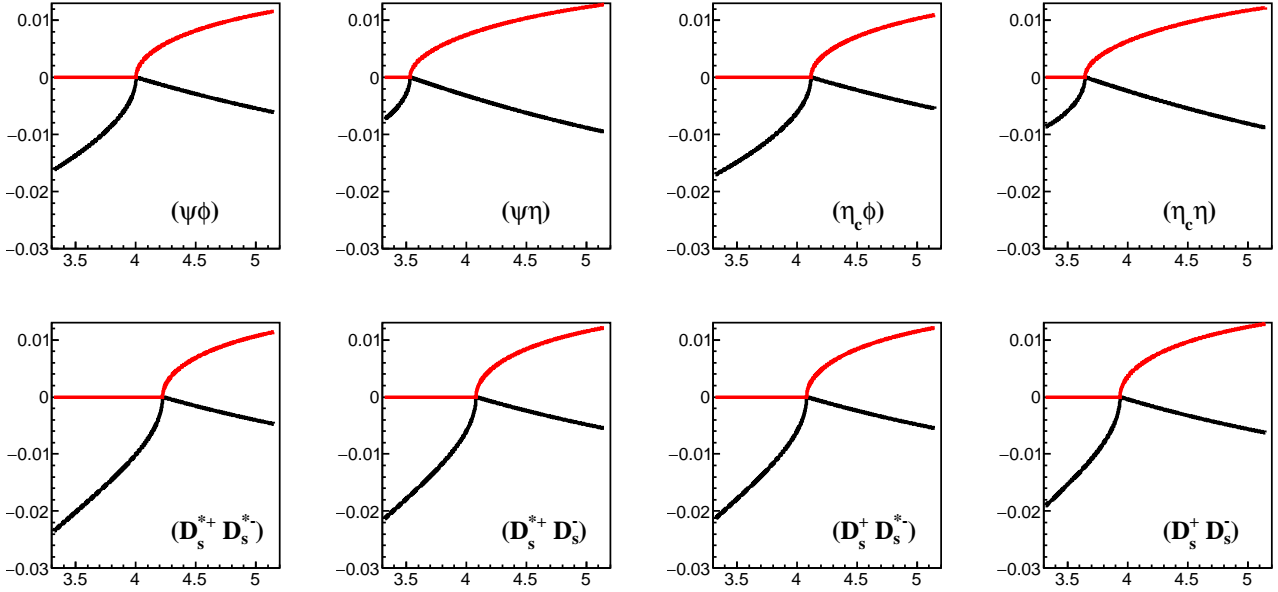,height=0.50\textwidth} }
\caption{ \label{f2}
Imaginary (red) and real (black) parts
of the loop diagrams $L_{\psi\phi}(s)$, $L_{\psi\eta}(s)$,
$L_{\eta_c\phi}(s)$, $L_{\eta_c\eta}(s)$,
$L_{D^{*+}_sD^{*-}_s}(s)$, $L_{D^{*+}_sD^-_s}(s), L_{D^{+}_sD^{*-}_s}(s)$,
$L_{D^+_sD^-_s}(s)$ at $g^2=1$ GeV$^2$ as a functions of energy, $\sqrt s$.
Thresholds are singular points.
 }
 \end{figure}

\subsection{Loop diagrams for meson-meson rescatterings }

Here for the production amplitudes of the studied diquark-antidiquark resonances
($2^{++},$ $1^{+-},$ $0^{++};$ $1^{++},$ $1^{+-};$ $0^{++}$) we present decomposition of
the spin wave functions into meson-meson ones and calculate corresponding meson-meson
loop diagrams. In this way we write the D-functions which contain poles inherent to the
resonances.

\subsubsection{One-pole amplitude for the \boldmath $2^{++}$ - state}

Diquark-antidiquark classification gives one level for the S-wave
$2^{++}$ - state that results in
one-pole production amplitude.  The spin wave function convolution
(see Appendix B for detail)
is  written for ($J_z=0$) - state as follows:
\bea \label{12}
&&
W^{(2^{++})}_{(A_{(cs)}A_{(\bar c\bar s)})\cdot (A_{(cs)}A_{(\bar c\bar s)})}\ =\
\Braket{\Psi^{(2^{++},J_z=0)}_{(A_{(cs)}A_{(\bar c\bar s)})}|
        \Psi^{(2^{++}),J_z=0}_{(A_{(cs)}A_{(\bar c\bar s)})} }
\\ \nn
&&
=\frac 23\braket{\psi^{(0)}\phi^{(0)}|\psi^{(0)}\phi^{(0)}}+
\frac
16\braket{\psi^{(\Uparrow)}\phi^{(\Downarrow)}|\psi^{(\Uparrow)}\phi^{(\Downarrow)}} +
\frac 16\braket{\psi^{(\Downarrow)}\phi^{(\Uparrow)}|\psi^{(\Downarrow)}\phi^{(\Uparrow)}
}
\\ \nn
&&
+\frac23\braket{D^{*+(0)}_s\,D^{*-(0)}_s|D^{*+(0)}_s\,D^{*-(0)}_s}+
\frac16\braket{D^{*+(\Uparrow)}_s\, D^{*-(\Downarrow)}_s|
D^{*+(\Uparrow)}_s\,D^{*-(\Downarrow)}_s}
\\ \nn
&&
+
\frac16\braket{D^{*(\Downarrow)}_s\,D^{*-(\Uparrow)}_s|
D^{*(\Downarrow)}_s\,D^{*-(\Uparrow)}_s }.
\eea
Loop diagrams $L_{2^{++}}$ (see Appendix A) are equal to:
\bea \label{13}
&&
L_{\psi\phi}(s)=
\int\limits_{(M_\psi+M_\phi)^2}^{+\infty}\frac{ds'}{\pi}
\frac{\rho_{\psi\phi}(s') }{s'-s-i0}\,,
\qquad
L_{D^{*+}_s D^{*-}_s}(s)=
\int\limits_{4M^2_{D^{*\pm}_s}}^{+\infty}\frac{ds'}{\pi}
\frac{\rho_{D^{*+}_s D^{*-}_s}(s') }{s'-s-i0}\,,
\eea
where phase space factor reads
\be  \label{14}
\rho_{\psi\phi}(s)=\frac{\sqrt{[s-(M_\psi+M_\phi)^2][s-(M_\psi-M_\phi)^2]}}{16\pi s}
\;\Theta\left(s-(M_\psi+M_\phi)^2\right)\,
\ee
with $\Theta(x<0)=0$, $\Theta(x>0)=1$. For the $\rho_{D^{*+}_sD^{*-}_s}(s)$ one should
replace in (\ref{14}):
 $M_\psi\to M_{D^{*+}_s} $ and  $M_\phi\to M_{D^{*-}_s} $.

The loop diagrams used in calculations are shown in Fig. {\ref{f2}, see also Appendix A.
We keep $g^2=const$ restoring the convergence of the dispersion relation integrals by the subtraction procedure.

Resonance production amplitude in D-function technique (for example, see \cite{book3})
reads:
\be
\label{15}
A^{(2^{++})}_{X\to \alpha}=g_X\frac{1}{m^2_{2^{++}}-s-g^2
\bigg[ L_{\psi\phi}(s)+L_{D^{*+}_s D^{*-}_s}(s)\bigg]}  g_{\alpha}\,,
\quad \alpha\ =\ \psi\phi,\,  D^{*+}_s D^{*-}_s \,,
\ee
where  $g_X, g_\alpha$ refer to initial and final state couplings and
$m_{2^{++}}$ is the input resonance  mass
which is roughly estimated in eq. (\ref{11}): $m_{2^{++}}\simeq (4580-4680)$ MeV.
Recall that the convergency of integrals with constant couplings  $g_X, g_\alpha$ is ensured by
subtraction procedure, see Appendix A for details.

Eq.(\ref{15}) presents a generalization of the Breit-Wigner equation, graphically it is
shown in Fig. \ref{f3}  as an infinite set of the meson-meson loop diagrams.

\begin{figure}[h!]
\centerline{\epsfig{file=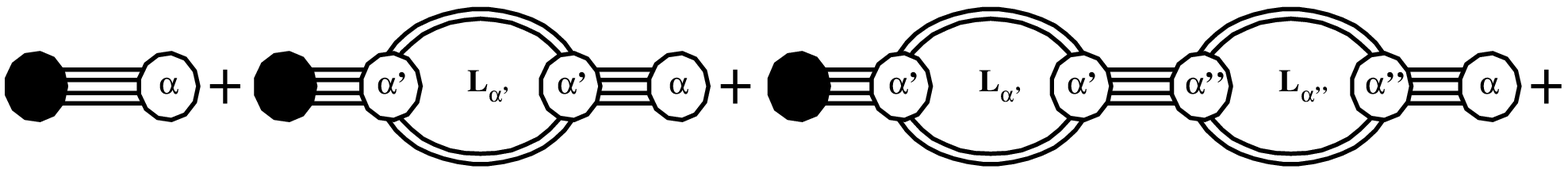,height=0.15\textwidth}}
\centerline{\epsfig{file=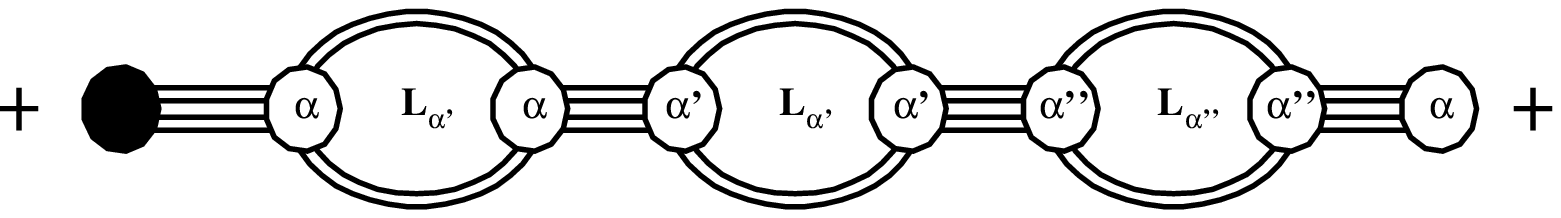,height=0.15\textwidth}
}
\caption{ \label{f3}
D-function pole amplitude as infinite set of diagrams with meson-meson rescatterings.
For the $2^{++}$ state summing is performed over
$\alpha, \alpha', \alpha''= \psi\phi,\,  D^{*+}_s
D^{*-}_s$.} \end{figure}

\begin{figure}
\centerline{\epsfig{file=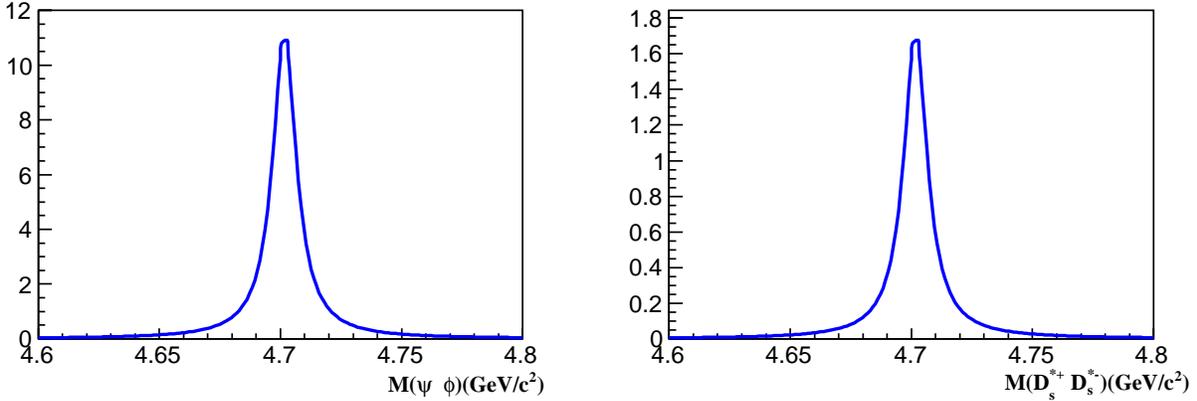,height=0.35\textwidth}
}
\caption{ \label{f4}
Production densities of the $\psi\phi$ and $D^{*+}D^{*-}$ states in the $2^{++}$
spectrum.
}
\end{figure}

The loop diagram $L_{D^{*+}_sD^{*-}_s}(s)$
 reads:
\bea \label{20}
&&L_{D^{*+}_sD^{*-}_s}(s)=
 \frac{1}{\pi}
\sqrt{
\frac{s-4M^2_{D_s^{*\pm}}}{s}
}
\bigg[\frac{1}{\pi}\ln
\frac{\sqrt{s}-\sqrt{s-4M_{D_s^{*\pm}}^2}}
{\sqrt{s}+\sqrt{s-4M_{D_s^{*\pm}}^2}}
 +i\bigg]\,,    \qquad
s>4M_{D^{*\pm}_s}^2
\\
&&L_{D^{*+}_sD^{*-}_s}(s)=
 \frac{i}{\pi}
\sqrt{
\frac{-s+4M^2_{D_s^{*\pm}}}{s}
}
\bigg[-\frac{2i}{\pi}\, \tan^{-1}\bigg(
\frac{
\sqrt{-s+4M_{D_s^{*\pm}}^2}
}
{\sqrt{s}}\bigg)
 +i\bigg]\,,    \qquad
s<4M_{D^{*\pm}_s}^2 \,.
\nn
 \eea
Subtraction is chosen to have  $L_{D^{*+}_sD^{*-}_s}(s=4M^2_{D_s^{*\pm}})=0$.

Separating real and imaginary parts of the amplitude, $A=\Re A+i\Im A$, we write
for the particle production densities:
\bea  \label{21}
\nn
&&
\sum_{S_z}\left|A^{(2^{++})}_{X\to\psi\phi}\right|^2 \rho_{\psi\phi} =
 \frac{g^2_X\, g^2_{\alpha}\,\rho_{\psi\phi} }
{\Big[m^2_{2^{++}}-s
-g^2\,  \left({\bf \Re}L_{\psi\phi}+{\bf \Re}L_{D^{*+}_s D^{*-}_s}\right)\Big]^2 +
\Big[
 g^2\, \left({\bf \Im}L_{\psi\phi}+{\bf \Im}L_{D^{*+}_s D^{*-}_s}\right)\Big]^2} \,,
\\ \nn
&&
\sum_{S_z}\left|A^{(2^{++})}_{X\to D^{*+}_s D^{*-}_s}\right|^2 \rho_{D^{*+}_s D^{*-}_s}
\\
&&
=\ \frac{g^2_X\, g^2_\alpha\,\rho_{D^{*+}_s D^{*-}_s}}
{\Big[m^2_{2^{++}}-s -g^2\,  \left({\bf
\Re}L_{\psi\phi}+{\bf \Re}L_{D^{*+}_s D^{*-}_s}\right)\Big]^2 + \Big[
 g^2\, \left({\bf \Im}L_{\psi\phi}+{\bf \Im}L_{D^{*+}_s D^{*-}_s}\right)\Big]^2}\,,
\eea
where summing is performed over meson spins at fixed $J_z$, see Eq. (\ref{12}).
Let us emphasize that the particle production densities are
equal zero below the corresponding thresholds, namely,
at $s<(M_\psi+M_\phi)^2 $ for the $(\psi\phi)$-channel
and  at $s<4M^2_{D_s^{*\pm}}$
for the $(D^{*+}_s D^{*-}_s)$-channel.
Production densities for the $2^{++}$-state are shown in Fig. \ref{f4}.

\begin{figure}
\centerline{\epsfig{file=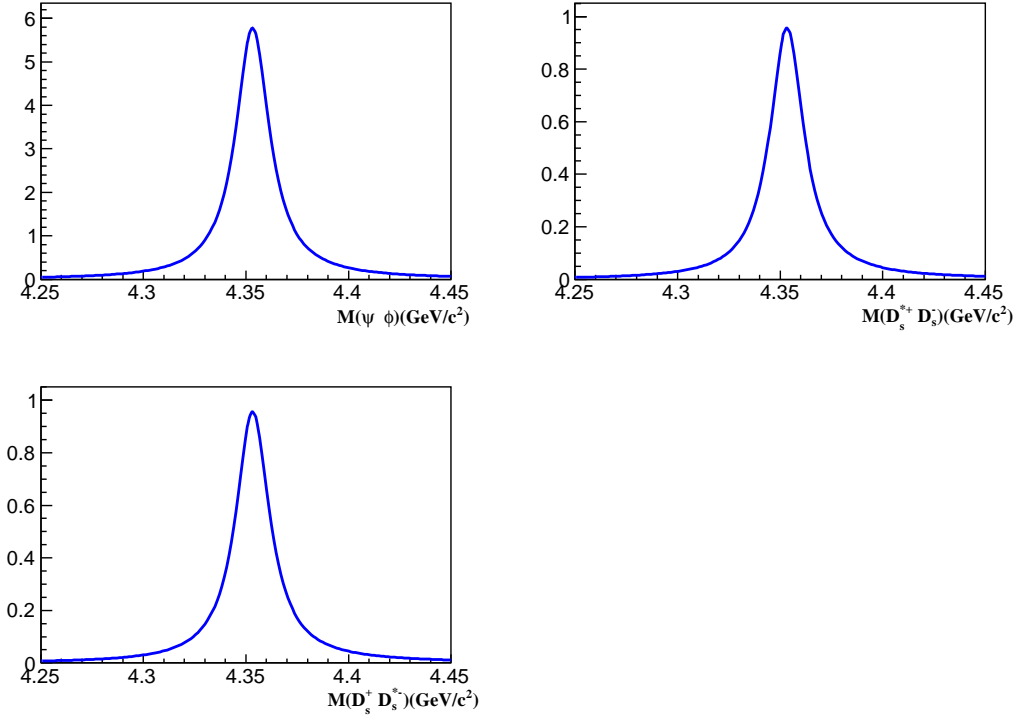,height=0.6\textwidth}
}
\caption{ \label{f5}
Production densities in $\psi\phi$ and $D^{+}D^{*-}$, $D^{*+}D^{-}$ spectra for the
$1^{++}$-states. }
\end{figure}

\subsubsection{One-pole amplitude for the \boldmath $1^{++}$ - state}

For the $(J^{PC}=1^{++},J_z=0)$ - state the wave function
convolution in meson space reads:
\bea
W^{(1^{++})}_{[S_{(cs)}A_{(\bar c\bar s)}]}& =&
 \frac12\braket{\psi^{(\Uparrow)}\phi^{(\Downarrow)}|\psi^{(\Uparrow)}\phi^{(\Downarrow)}}
+\frac12\braket{\psi^{(\Downarrow)}\phi^{(\Uparrow)}|\psi^{(\Downarrow)}\phi^{(\Uparrow)}}
\\ \nn
&&
+\frac12\braket{D^{*+(0)}_s\,D^{-}_s|D^{*+(0)}_s\,D^{-}_s}
+\frac12\braket{D^{+}_s\,D^{*-(0)}_s|D^{+}_s\,D^{*-(0)}_s}\,,
\eea
and the amplitude is written as:
\bea
A^{(1^{++})}_{X\to \alpha}&=&
g_X\,\frac{1}{m^2_{1^{++}}-s-g^2\,
\Big[L_{\psi\phi}(s)+\frac12 L_{D^{*+}_s\,D^{-}_s}(s)+\frac12 L_{D^{+}_s\,D^{*-}_s}(s)\Big]}\,
g_{\alpha}\,,
\\ \nn
&&\alpha=  \psi\phi, D^{*+}_s\,D^{-}_s, D^{+}_s\,D^{*-}_s\,.
\eea
with
bare mass  given in eq. (\ref{11}),
$m_{1^{++}}\simeq (4310-4350)$ MeV.

Production densities for the $1^{++}$ - state are shown in Fig. \ref{f5}.

\begin{figure}[h!]
\centerline{\epsfig{file=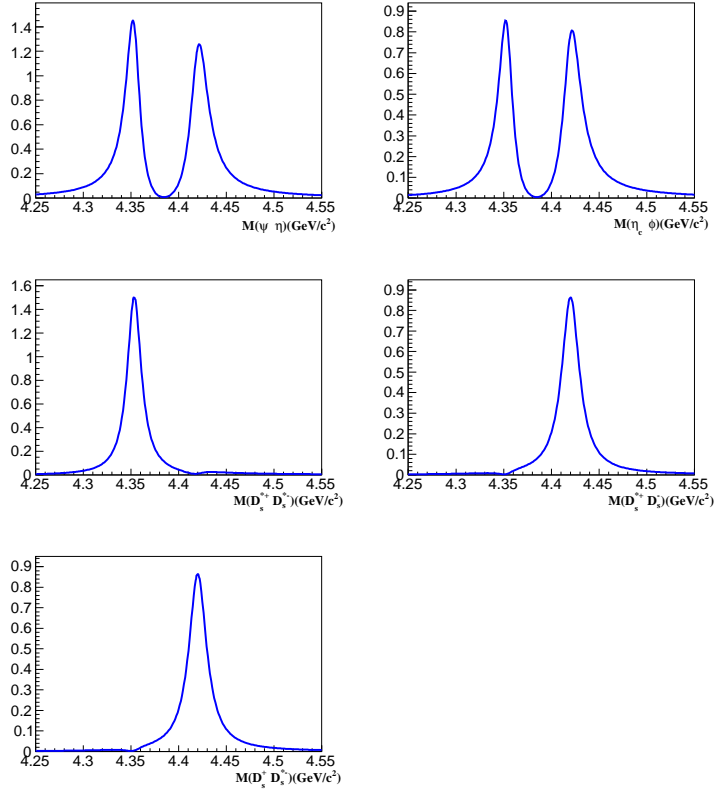,height=0.65\textwidth}
}
\caption{ \label{f6}
Production densities for the $1^{+-}$ - states. The red and green cures
refer to case with switched-off mixing of different states.}
\end{figure}

\subsubsection{ Two-pole amplitude for the \boldmath$1^{+-}$ - states}

For the $J^{PC}=1^{+-}$ we have two levels
 which can recombinate into three meson-meson states.
The wave functions for the $(J=1,J_z=0)$ - states read:
\bea
\Psi^{(1^{+-},J_z=0)}_{(A_{(cs)}A_{(\bar c\bar s)})}&=&\frac{1}{\sqrt{2}}
\Big[\psi^{(0)}\eta+\eta_c \phi^{(0)}-D^{*+(0)}_s\, D^-_s-D^+_s\, D^{*-(0)}_s\Big]\,,
\\
\Psi^{(1^{+-},J_z=0)}_{\{S_{(cs)}A_{(\bar c\bar s)}\}}&=&
 \frac{1}{\sqrt{2}}\Big[
-\psi^{(0)}\eta+\eta_c \phi^{(0)}
-D^{*+(\Uparrow)}_s D^{*-(\Downarrow)}_s
+D^{*+(\Downarrow)}_s D^{*-(\Uparrow)}_s
\Big]\,.
\nn
\eea

Correspondingly we write the three meson-meson convolutions:
\bea
&&
W^{(1^{+-})}_{(A_{(cs)}A_{(\bar c\bar s)})} =
\Braket{\Psi^{(1^{+-},J_z=0)}_{(A_{(cs)}A_{(\bar c\bar s)})}|
        \Psi^{(1^{+-},J_z=0)}_{(A_{(cs)}A_{(\bar c\bar s)})}}
\ =\frac{1}{2}\Braket{\psi^{(0)}\eta|\psi^{(0)}\eta}
\\ \nn
&&
+\frac{1}{2}\Braket{\eta_c \phi^{(0)}|\eta_c \phi^{(0)}}
+\frac12\Braket{ D^{*+(0)}_s D^{-}_s|D^{*+(0)}_s D^{-}_s}+
 \frac12\Braket{ D^{+}_s D^{*-(0)}_s|D^{+}_s D^{*-(0)}_s},
\\ \nn
&&
W^{(1^{+-})}_{(A_{(cs)}A_{(\bar c\bar s)})\cdot \{S_{(cs)}A_{(\bar c\bar s)}\}}\ =\
\Braket{\Psi^{(1^{+-,J_z=0})}_{(A_{(cs)}A_{(\bar c\bar s)})}|
        \Psi^{(1^{+-},J_z=0)}_{\{S_{(cs)}A_{(\bar c\bar s)}\}}}
\\ \nn
&&
\ =\ -\frac{1}{2}\Braket{\psi^{(0)}\eta|\psi^{(0)}\eta}
+\frac{1}{2}\Braket{\eta_c \phi^{(0)}|\eta_c \phi^{(0)}}
\\ \nn
&&
W^{(1^{+-})}_{\{S_{(cs)}A_{(\bar c\bar s)}\}}
=
\Braket{\Psi^{(1^{+-},J_z=0)}_{\{S_{(cs)}A_{(\bar c\bar s)}\}}|
\Psi^{(1^{+-},J_z=0)}_{\{S_{(cs)}A_{(\bar c\bar s)}\}}}\ =\
 \frac{1}{2}\Braket{ \psi^{(0)}\eta|\psi^{(0)}\eta }
+\frac{1}{2}\Braket{ \eta_c \phi^{(0)}|\eta_c \phi^{(0)}  }
\nn
\\
&&
+\frac{1}{2}\Braket{ D^{*+(\Uparrow)}_s\, D^{*-(\Downarrow)}_s|D^{*+(\Uparrow)}_s\,
D^{*-(\Downarrow)}_s }
+\frac{1}{2}\Braket{ D^{*+(\Downarrow)}_s\, D^{*-(\Uparrow)}_s|D^{*+(\Downarrow)}_s\,
D^{*-(\Uparrow)}_s } ,
\nn
\eea
and
three loop diagram combinations:
\bea
L^{(1^{+-})}_{11}&=&\frac12 g^2\,L_{\eta_c \phi}
+\frac12 g^2\,L_{\psi\eta}+g^2\,\frac12 L_{D^{*+}_s D^{-}_s}
+g^2\,\frac12 L_{D^{+}_s D^{-*}_s}\,,
\\
L^{(1^{+-})}_{12}&=&
-\frac1{2} g^2\,L_{\psi\eta}
+\frac1{2} g^2\,L_{\eta_c \phi}\,\,,
\nn
\\
L^{(1^{+-})}_{22}&=&\frac12 g^2\,L_{\eta_c \phi}+\frac12 g^2\,L_{\psi\eta}
+ g^2\,L_{D^{*+}_s D^{*-}_s}\,.
\nn
\eea
Indices 1, 2 refer to levels  $1\equiv \{A_{(Qs)}A_{(\bar Q\bar s)}\}$ and
$2\equiv \{S_{(Qs)}A_{(\bar Q\bar s)}\}$.
The loop diagram $L^{(1^{+-})}_{12}$ describes non-diagonal transition
$(A_{(Qs)}A_{(\bar Q\bar s)})\to  \{S_{(Qs)}A_{(\bar Q\bar s)}\}$,
it is relatively small due to the partial cancellation of two contributions:
$L^{(1^{+-})}_{12}=\frac1{2} g^2\,\Big (-L_{\psi\eta}+L_{\eta_c \phi}
\Big)$
.

In the general case the amplitude reads:
\bea
A^{(1^{+-})}_{(X\to \alpha)}
&=&
g_{(X\to 1)}\frac{1}{\Delta^{(1^{+-})}}
\bigg[d^{(1^{+-})}_{1}(1-L^{(1^{+-})}_{22}d^{(1^{+-})}_2)
+d^{(1^{+-})}_2L^{(1^{+-})}_{21}d^{(1^{+-})}_1\bigg]g_{(1\to\alpha)}
\\
&&+
g_{(X\to 2)}\frac{1}{\Delta^{(1^{+-})}}
\bigg[d^{(1^{+-})}_{2}(1-L^{(1^{+-})}_{11}d^{(1^{+-})}_1)
+d^{(1^{+-})}_1L^{(1^{+-})}_{12}d^{(1^{+-})}_2\bigg]g_{(2\to\alpha)}
\nn \\
&=&
g_{(X\to 1)}\frac{1}{\bigtriangledown^{(1^{+-})}}
\bigg[m^2_{1^{+-}\{AS\}}\,-s-L^{(1^{+-})}_{22}
+L^{(1^{+-})}_{21}\bigg]g_{(1\to\alpha)}
\nn \\
&&
+
g_{(X\to 2)}\frac{1}{\bigtriangledown^{(1^{+-})}}
\bigg[m^2_{1^{+-}(AA)}\,-s-L^{(1^{+-})}_{11}
+L^{(1^{+-})}_{12}\bigg]g_{(2\to\alpha)} \,,
\nn
\eea
where
\bea
\label{bigtriangledown}
&&
\Delta^{(1^{+-})}=(1- L^{(1^{+-})}_{11}d^{(1^{+-})}_1)(1-L^{(1^{+-})}_{22}d^{(1^{+-})}_2)
-L^{(1^{+-})}_{12}d^{(1^{+-})}_2 L^{(1^{+-})}_{21}d^{(1^{+-})}_1 \,,
\\
&&
\bigtriangledown^{(1^{+-})}=\Big(m^2_{1^{+-}(AA)}\,-s- L^{(1^{+-})}_{11}\Big)
\Big(m^2_{1^{+-}\{AS\}}\,-s-L^{(1^{+-})}_{22}\Big)
-L^{(1^{+-})}_{12}L^{(1^{+-})}_{21} \,.
\nn
\eea
Bare masses $m_{1^{+-}(AA)}\simeq (4380-4420)$ MeV and
$m_{1^{+-}\{AS\}}\simeq (4310-4350)$ MeV are given in eq. (\ref{11}).

Zeros of the $\bigtriangledown$ (eq.(\ref{bigtriangledown})) on the complex-$s$ plane determine masses and widths of two resonances.
Meson production densities for $1^{+-}$ - states are demonstrated in Fig. \ref{f6}.

\begin{figure}
\centerline{\epsfig{file=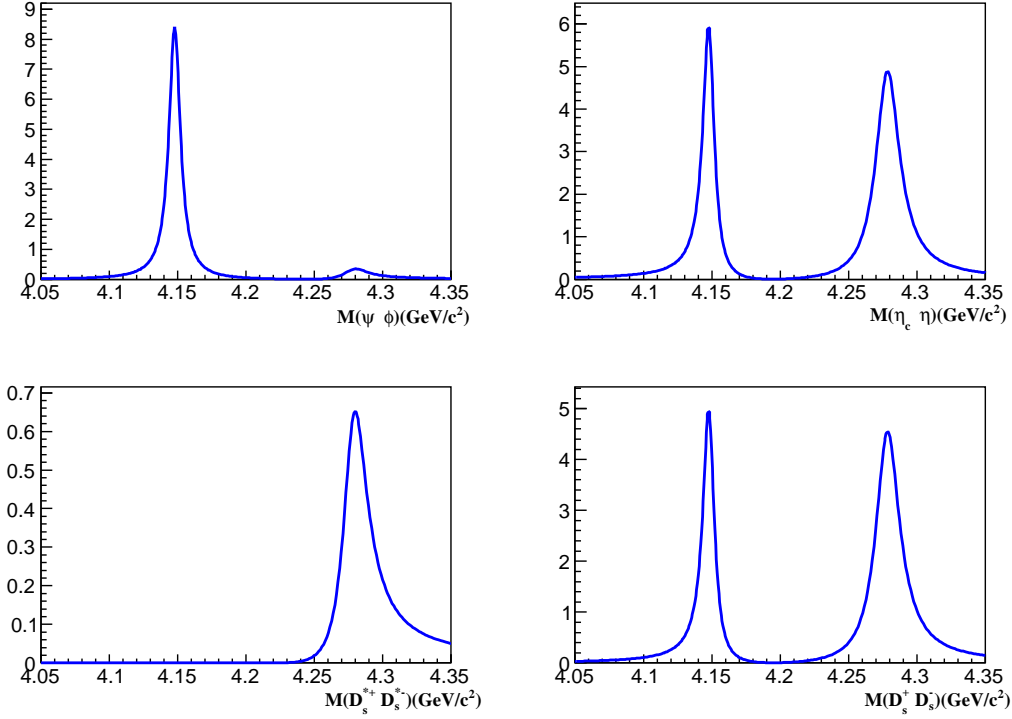,height=0.6\textwidth}
}
\caption{ \label{f7}
Production densities for the $0^{++}$-states.}
\end{figure}

\subsubsection{ Two-pole amplitude for the \boldmath$0^{++}$ - state}

The wave functions for scalar states  read:
\bea
&&\Psi^{(0^{++})}_{(A_{(cs)}A_{(\bar c\bar s)})}\ =\
\frac{1}{2\sqrt{3}}\Big[
\psi^{(0)}\phi^{(0)}-\psi^{(\Uparrow)}\phi^{(\Downarrow)}
-\psi^{(\Downarrow)} \phi^{(\Uparrow)}+3\eta_c\eta
\\ \nn
&&-D^{*+(0)}_s\,D_s^{*-(0)}
+D^{*+(\Uparrow)}_s\, D_s^{*-(\Downarrow)}
+D^{*+(\Downarrow)}_s\, D_s^{*-(\Uparrow)}
-3D^{+}_s D^{-}_s
\Big]
\\ \nn
&&
\Psi^{(0^{++})}_{(S_{(cs)}S_{(\bar c\bar s)})}\ =\
\frac{1}{2}\Big[\psi^{(\Uparrow)} \phi^{(\Downarrow)}
+\psi^{(\Downarrow)}\phi^{(\Uparrow)}
- \psi^{(0)}\phi^{(0)}+\eta_c\eta
\\ \nn
&&- D^{*+(0)}_s\, D^{*-(0)}_s
+D^{*+(\Uparrow)}_s\, D^{*-(\Downarrow)}_s
+D^{*+(\Downarrow)}_s\, D^{*-(\Uparrow)}_s
+D^{+}_s\, D^{-}_s\Big].
\eea
For two $0^{++}$ levels one has three wave function convolutions:
 \bea
&&
\Braket{\Psi^{(0^{++})}_{(A_{(cs)}A_{(\bar c\bar s)})}|
        \Psi^{(0^{++})}_{(A_{(cs)}A_{(\bar c\bar s)})}}\ =\
\\
&&
\frac{1}{4}\Braket{\psi\phi|\psi\phi}
+\frac34\Braket{\eta_c\eta|\eta_c\eta}
+\frac14\Braket{D^{*+}_s D^{*-}_s|D^{*+}_s D^{*-}_s}
+\frac34\Braket{D^{+}_s D^{-}_s|D^{+}_s D^{-}_s} \,,
\nn
\\
&&
\Braket{\Psi^{(0^{++})}_{(A_{(Qs)}A_{(\bar Q\bar s)})}|
        \Psi^{(0^{++})}_{(S_{(Qs)}S_{(\bar Q\bar s)})}}\ =\
\nn
\\
\nn
&&
-\frac{\sqrt 3}{4}\Braket{\psi\phi|\psi\phi}
+\frac{\sqrt 3}{4}\Braket{\eta_c \eta|\eta_c \eta}
+\frac{\sqrt 3}{4}\Braket{D^{*+}_s D^{*-}_s|D^{*+}_s D^{*-}_s}
-\frac{\sqrt 3}{4}\Braket{D^{+}_s D^{-}_s|D^{+}_s D^{-}_s} \,,
\\
&&
\Braket{\Psi^{(0^{++})}_{(S_{(Qs)}S_{(\bar Q\bar s)})}|
        \Psi^{(0^{++})}_{(S_{(Qs)}S_{(\bar Q\bar s)})}}\ =\
\nn
\\
\nn
&&
\frac{3}{4}\Braket{\psi\phi|\psi\phi}
+\frac{1}{4}\Braket{\eta_c \eta|\eta_c \eta}
+\frac{3}{4}\Braket{D^{*+}_s D^{*-}_s | D^{*+}_s D^{*-}_s}
+\frac{1}{4}\Braket{D^{+}_s D^{-}_s| D^{+}_s D^{-}_s}  \,,
\eea
and three transition
diagrams:
\bea
L^{(0^{++})}_{11}&=&\frac14 g^2\,L_{\psi\phi}+\frac34 g^2\,L_{\eta_c \eta}
+\frac14 g^2\,L_{D^{*+}_s D^{*-}_s}+\frac34 g^2\,L_{D^{+}_s D^{-}_s}
\,,\\
\nn
L^{(0^{++})}_{12}&=&-\frac{\sqrt3}4 g^2\,L_{\psi\phi}+\frac{\sqrt3}4 g^2\,L_{\eta_c
\eta} +\frac{\sqrt3}4 g^2\,L_{D^{*+}_s D^{*-}_s}-\frac{\sqrt3}4 g^2\,L_{D^{+}_s D^{-}_s}
\,,\\
\nn
L^{(0^{++})}_{22}&=&\frac34 g^2\,L_{\psi\phi}+\frac14 g^2\,L_{\eta_c\eta}
+\frac34 g^2\,L_{D^{*+}_s D^{*-}_s}+\frac14 g^2\,L_{D^{+}_s D^{-}_s}
\,.
\eea

The amplitude reads:
\bea
A^{(0^{++})}_{(X\to \alpha)}
&=&
g_{X_1}\frac{1}{\Delta^{(0^{++})}}
\bigg[d^{(0^{++})}_{1}(1-L^{(0^{++})}_{22}d^{(0^{++})}_2)
+d^{(0^{++})}_2L^{(0^{++})}_{21}d^{(0^{++})}_1\bigg]g_{(1\to\alpha)}
\\
&&
+
g_{X_2}\frac{1}{\Delta^{(0^{++})}}
\bigg[d^{(0^{++})}_{2}(1-L^{(0^{++})}_{11}d^{(0^{++})}_1)
+d^{(0^{++})}_1L^{(0^{++})}_{12}d^{(0^{++})}_2\bigg]g_{(2\to\alpha)}
\nn
\\
&=&
g_{X_1}\frac{1}{\bigtriangledown^{(0^{++})}}
\bigg[m^2_{0^{++}(SS)}\,-s-L^{(0^{++})}_{22}
+L^{(0^{++})}_{21}\bigg]g_{(1\to \alpha)}
\nn
\\
&&
+
g_{X_2}\frac{1}{\bigtriangledown^{(0^{++})}}
\bigg[m^2_{0^{++}(AA)}\,-s-L^{(0^{++})}_{11}
+L^{(0^{++})}_{12}\bigg]g_{(2\to\alpha)} \,,
\nn
\eea
where
\bea
&&
\Delta^{(0^{++})}=\Big(1-L^{(0^{++})}_{11}d^{(0^{++})}_1\Big)
\Big(1-L^{(0^{++})}_{22}d^{(0^{++})}_2\Big)
-L^{(0^{++})}_{12}d^{(0^{++})}_2 L^{(0^{++})}_{21}d^{(0^{++})}_1 \,,
\nn\\
&&
\bigtriangledown^{(0^{++})}=\Big(m^2_{0^{++}(AA)}\,-s-L^{(0^{++})}_{11}\Big)
\Big(m^2_{0^{++}(SS)}\,-s -L^{(0^{++})}_{22}\Big)
-L^{(0^{++})}_{12}L^{(0^{++})}_{21} \,.
\nn
\eea
Production densities for the $1^{++}$ - state are shown in Fig. \ref{f7}.

\begin{figure}[h!]
\centerline{\epsfig{file=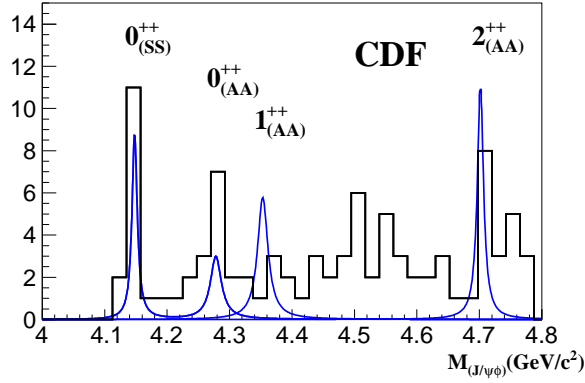,height=0.33\textwidth}
           }
\caption{ \label{f8}
The CDF-spectrum \cite{cdf-ss} for $ \phi J/\psi$ state in decay
$B^\pm \to\phi(J/\psi )K^\pm $ and its comparision with the diquark-antidiquark model:
$M_{0^{++}}\simeq 4140$ MeV ,
$M_{0^{++}}\simeq 4276$ MeV ,
 $M_{1^{++}}\simeq 4278$ MeV ,
$M_{2^{++}}\simeq 4500$ MeV , the parameters are as follows: $m_S=2072$ MeV,
$m_A=2137$ MeV, and $\Delta=70$ MeV.
}
\end{figure}

\begin{figure}
\centerline
{\epsfig{file=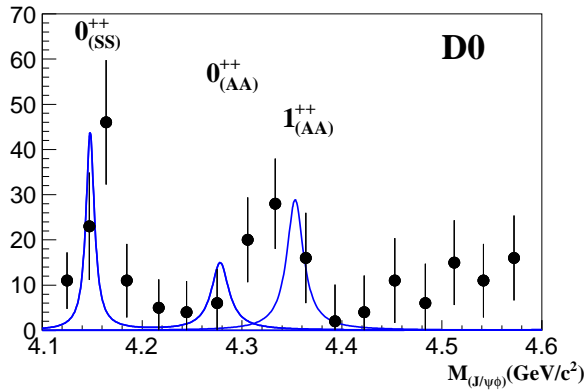,height=0.33\textwidth} }
\caption{ \label{f9} The D0-spectrum \cite{D0-ss} for $(J/\psi\phi )$ state
in decay $B^+\to (\phi J/\psi)K^+$.}
 \end{figure}

\begin{figure}[h!]
\centerline{\epsfig{file=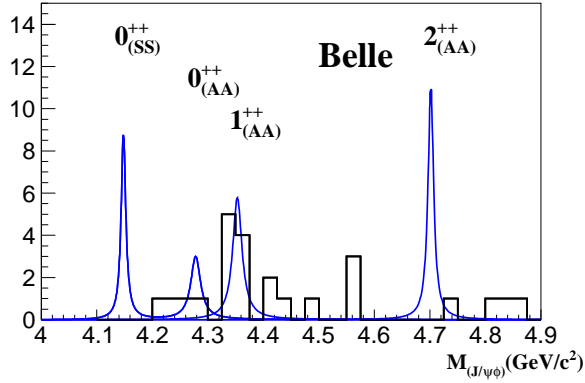,height=0.33\textwidth}}
\caption{ \label{f10}
The Belle data \cite{belle-ss} for reaction $\gamma\gamma\to \phi J/\psi$.}
\end{figure}

\subsubsection{Meson-meson components and shifts of  the resonance masses }

The meson-meson component shifts pole positions. For different states
they are as follows:
\bea
\label{mass-shift}
&& \delta m(0_{(AA)}^{++})=173 \mbox{
MeV},\qquad \delta m(0_{(SS)}^{++})=142 \mbox{ MeV}, \\
&& \delta m(1_{(AS)}^{++})=153
\mbox{MeV}, \nn\\
&& \delta m(1_{(AA)}^{+-})=111\mbox{ MeV}, \qquad \delta
m(1_{(AS)}^{+-})=68  \mbox{ MeV}, \nn\\
&& \delta m(2_{(AA)}^{++})=144 \mbox{ MeV}.
\nn
\eea
The considered states are not molecular-like or deutron-like systems  that affected
in hadronic values for the sifts of the resonance masses.

\section{ Data for ($\phi J/\psi$)-spectra and diquark-antidiquark states}

In Figs. \ref{f8} - \ref{f10} we demonstrate ($J/\psi$) - spectra measured
by  Belle \cite{belle-ss}, CDF \cite{cdf-ss} and D0 \cite{D0-ss}.
Using the classification of the diquark-antidiquark states suggested
in Section \ref{s222}
we compare calculated in Section \ref{sec3} production densities with experimental data.
The comparison allows to guess that eq. (\ref{11}) leads to correct systematics
of heavy exotic charmonia.

Fig. \ref{f10} shows that reaction of
$\gamma^*\gamma^*\to X^{(0^{++})}_{(s\bar s)}(4140)\to\phi J/\psi$ is
relatively suppressed. Indeed, within
vector dominance model,
$\gamma^*\to J/\psi\to (c\bar c)$ and $\gamma^*\to \phi\to (s\bar s)$,
we have for production ratio
$R(\frac{SS(0^{++})\to\phi\psi}{AS(1^{++})\to\phi\psi})\sim\frac14$. This factor
should be introduced additionally into production density curve in Fig. \ref{f10}.

\section{Exotic mesons with hidden strangeness as members of the nonet classification}

 Nonet classification of mesons (${\bf 1}+{\bf 8}$) gives a standard and successful way for
 treating the $q\bar q$ states ($q=u,d,s$). It is reasonable to suggest that a similar
 classification works when constituents are diquarks. Then, considering the system of three
 diquarks, $\Big((u\bar c),(d\bar c),(s\bar c)\Big)$ as an analog of the three-quark system
 $(u,d,s)$, we write six nonets for S-wave diquark-antidiquark systems:
\be \label{33}
\begin{tabular}{l|l|l|l|l|l}
& &$(cq)\cdot(\bar c\bar q)$ &$(cq)\cdot(\bar c\bar q)$
 &$(cs)\cdot(\bar c\bar s)$ & $(cq)\cdot(\bar c\bar s)$   \\
& &I=1 &I=0 &I=0 & $ I=1/2$   \\
\hline
$2^{++}$& $( A_{(cs)}\cdot A_{(\bar c\bar s)})$&$\sim$ 4430&$\sim$ 4430 &4630$\pm$ 50&$\sim$ 4530
\\
$1^{+-}$& $( A_{(cs)}\cdot A_{(\bar c\bar s)})$&$\sim$ 4200&$\sim$ 4200 &4400$\pm$ 50&$\sim$ 4300
\\
$0^{++}$& $( A_{(cs)}\cdot A_{(\bar c\bar s)})$&$\sim$ 4080&$\sim$ 4080 &4280$\pm$ 20&$\sim$ 4180
\\
 \hline
$1^{++}$& $[A_{(cs)}\cdot S_{(\bar c\bar s)}]$&$\sim$ 4180&$\sim$ 4180 &4380$\pm$ 20&$\sim$ 4280    \\
$1^{+-}$& $\{ A_{(cs)}\cdot S_{(\bar c\bar s)}\}$&$\sim$ 4130&$\sim$ 4130 &4330$\pm$ 50&$\sim$ 4230  \\
 \hline
$0^{++}$& $( S_{(cs)}\cdot S_{(\bar c\bar s)})$&$\sim$ 3930&$\sim$ 3930 &4140$\pm$ 20&$\sim$ 4040 \\
 \end{tabular}
\ee
Here we have used six diquark-antidiquark states with hidden strangeness,
$(cs)\cdot(\bar c\bar s)$, as a basis for the determination of masses of other nonet partners
(masses are in MeV units). Weighting of the strange quark is accepted to be
$\Delta m_s =100$ MeV, see discussion in Section 2, though the larger value, $\sim 130$
MeV, is possible.

Rich resonance structure in the exotic charmonium sector imply  a principal
concern in the study corresponding spectra: resonances in crossing channels or rescatterings
in direct channels can affect determination of resonance characteristics,
examples of such affections can be found in refs.
\cite{PRD50-1972-94,PRD51-R4619-22-95}.

Search for resonance states in the $(cq)\cdot(\bar c\bar q)$ sector was done in a series
of studies, see refs. \cite{babar-qq,belle-qq,LHCb-qq,cdf-qq,cleo-qq}. A set of
candidates for non-strange exotic states is discussed in \cite{PDG}.
Nevertheless, reliable comparison of eq. (\ref{33}) with data requires
much more  information.

\section{Conclusion}

We consider exotic mesons with hidden charm and strangeness in the mass region
(4100 - 4800) MeV
 as two-component composite systems with
(i) diquark-antidiquark component $(cs)\cdot(\bar c\bar s)$, and
(ii) meson-meson component $(c\bar s)\cdot(s\bar c)$. The notion of diquarks is actively
used in hadron physics both for mesons \cite{maiani,voloshin,ali} and baryons
\cite{qD,book4}. Following these ideas we construct a model in which
the meson-meson component is taken into account in addition. Supposing that the recombination
process $(cs)\cdot(\bar c\bar s)\to(c\bar s)\cdot(s\bar c)$ dominates we calculate
relative probabilities for decays into meson channels: $\psi\phi$ ,
$\eta_c\eta$ , $\eta_c\phi$ , $\psi\eta$ ,  $D^{*}_s\,\bar D^{*}_s$ ,
$D^{*}_s\,\bar D^{}_s$ ,
$D^{}_s\,\bar D^{*}_s$ ,
$D_s\bar D_s$.
Comparison with data \cite{belle-ss,cdf-ss,CMS-ss,D0-ss} is performed,
Figs. \ref{f8}, \ref{f9}, \ref{f10}.
Predictions for new states are presented, and
the nonet structure for $\Big((Qq)(\overline{Q q})\Big)$, \Big($(Qs)(\bar Q\bar s$)\Big),
\Big($(Qq)(\bar Q\bar s)\Big)$ - states ($q=u,d$) is suggested in eq. (\ref{33}).

\subsubsection*{Acknowledgment}

We thank A.K. Likhoded and J. Nyiri for  stimulating and useful disscutions.
 The work was supported by grants RSGSS-4801.2012.2.,  RFBR-13-02-00425 and RSCF-14-22-00281 .

\def\thesection{Appendix \Alph{section}}
\def\theequation{\Alph{section}.\arabic{equation}}
\setcounter{section}{0}

\section{Loop diagrams}
\setcounter{equation}{0}

We present loop diagrams for meson states calculated in terms of the dispersion relation
technique, for details see for example ref. \cite{book3}.

\subsection{Loop diagram for one-pole and one-channel amplitude}

\subsubsection*{Loop diagram above threshold, at $s>(M_a+M_b)^2$}

The equation for one-pole and one-channel D-function reads:
\bea
&&
D=d+D\,g^2L_{}\,d_ ,\\
&&
d=\frac{1}{m^2-s},\quad
L_{}=\int\limits_{(M_a+M_b)^2}^\infty\frac{ds'}{\pi}
\frac{\rho(s') }{s'-s-i0}.
\nn
\eea
Here
$m$ is a bare mass of this state, $g^2L_{}$ is loop diagram formed by hadrons,
$M_a,M_b$ are
masses of the loop mesons. The phase space factor reads:
\be
\rho_{ab}(s)=\frac{1}{16\pi s}\sqrt{[s-(M_a+M_b)^2][s-(M_a-M_b)^2]}\,.
\ee
At $s<4M^2$ we replace $\sqrt{s-4M^2}\to i\sqrt{4M^2-s}$, the point
$s=0$ is not singular for $L$.

The convergency of the integral for $L_{s}$ can be organized either due to
introducing a $s$-dependence of the vertex $g\to g(s)$ or by switching the subtraction
procedure:
\be \label{51}
L_{(ab)}(s)=\int\limits_{(M_a+M_b)^2}^\infty\frac{ds'}{\pi}
\cdot\frac{\rho(s')}{s'-s-i0}\to  \ell_0+
\int\limits_{(M_a+M_b)^2}^\infty
\frac{ds'}{\pi}\frac{s-(M_a+M_b)^2}{(s'-(M_a+M_b)^2)(s'-s-i0)}
 \rho(s')\,.
\ee

In the $(ab)$-channel we write for positive $s$, $s>(M_a+M_b)^2$:
\bea \label{52}
&& L_{(ab)}(s)=\ell_0+ \frac{\lambda}{s}
\\
\nn
&&
+\frac{\sqrt{[s-(M_a+M_b)^2][s-(M_a-M_b)^2]}}{16\pi s}
\bigg[\frac{1}{\pi}\ln
\frac{\sqrt{s-(M_a-M_b)^2}-\sqrt{s-(M_a+M_b)^2}}
{\sqrt{s-(M_a-M_b)^2}+\sqrt{s-(M_a+M_b)^2}}
 +i\bigg]\,,
\\
\nn
 &&\lambda=\frac{\sqrt{(M_a+M_b)^2(M_a-M_b)^2}}{16\pi^2}
 \ln\frac{\sqrt{(M_a+M_b)^2}+\sqrt{(M_a-M_b)^2}}
 {\sqrt{(M_a+M_b)^2}-\sqrt{(M_a-M_b)^2}}\quad .
 \eea
The point $s=(M_a+M_b)^2$ is singular.
For $s<(M_a+M_b)^2$ we write
$\sqrt{s-(M_a+M_b)^2}\to i\sqrt{(M_a+M_b)^2-s}$,
the points
$s=(M_a-M_b)^2$ and $s=0$
are not singular, the pole singularity at $s=0$
is canceled due to choice of $\lambda$. The subtraction constant $\ell_0$ is choosen to have
zero value for loop diagram at threshold
$L_{(ab)}(s)\Big|_{s=(M_a+M_b)^2}=0$, namely:
\be
\ell_0=\frac{-\lambda}{(M_a+M_b)^2}.
\ee

\subsubsection*{Loop diagram below threshold, at $s<(M_a+M_b)^2$}

At $(M_a-M_b)^2<s<(M_a+M_b)^2$ the loop diagram reads:
\bea
\label{19}
&&L_{(ab)}(s)=\ell_0+ \frac{\lambda}{s}
\\
\nn
&&
+i\frac{\sqrt{[-s+(M_a+M_b)^2][s-(M_a-M_b)^2]}}{16\pi s}
\bigg[\frac{1}{\pi}\ln
\frac{\sqrt{s-(M_a-M_b)^2}-i\sqrt{-s+(M_a+M_b)^2}}
{\sqrt{s-(M_a-M_b)^2}+i\sqrt{-s+(M_a+M_b)^2}}
 +i\bigg]
\\
&&
=\ell_0+
\frac{\lambda}{s}
+i\frac{\sqrt{[-s+(M_a+M_b)^2][s-(M_a-M_b)^2]}}{16\pi s}
\bigg[-\frac{2i}{\pi}\,\tan^{-1}\bigg(
\frac{\sqrt{-s+(M_a+M_b)^2}}{\sqrt{s-(M_a-M_b)^2}} \bigg)
+i \bigg] .
\nn
 \eea
The last line demonstrates the absence of a singularity in $s=(M_a-M_b)^2$.
Indeed, within top-down approaching to this point we have:
$$
-\frac{2i}{\pi}\,\tan^{-1}\bigg(
\frac{\sqrt{-s+(M_a+M_b)^2}}{\sqrt{s-(M_a-M_b)^2}} \bigg)+i=
-\frac{2i}{\pi}\bigg(
\frac{\pi}{2}- \tan^{-1}
\frac{\sqrt{s-(M_a-M_b)^2}}{\sqrt{-s+(M_a+M_b)^2}} \bigg)+i
$$
$$    \simeq
-\frac{2i}{\pi}\bigg(
\frac{\pi}{2}-
\frac{\sqrt{s-(M_a-M_b)^2}}{\sqrt{-s+(M_a+M_b)^2}} \bigg)+i
$$
with the corresponding cancellation of the singular terms in Eq. (\ref{19}).

\subsection{Multi-channel two-pole amplitude}

Now we consider a realistic situation, multi-channel amplitude.
For the one-pole case we write
\bea
&&
A_{(X\to \alpha)}\ =\
g_X\frac{d}{1-\sum\limits_{\alpha'}g_{\alpha'}^2L_{\alpha'}\,d} g_{\alpha}\,.
\\ &&
L_{\alpha'}=
\int\limits_{(M_a+M_b)^2}^{+\infty}\frac{ds'}{\pi}
\frac{\rho_{\alpha'}(s')}{s'-s-i0},
\nn
\\
&&
\alpha, \alpha'=
 \psi\phi,\, \eta_c\eta,\,\psi\eta,\, \eta_c\phi,
 \, D_s\bar D_s,\, D^*_s\bar D_s,\, D_s\bar D^*_s,\,
 D^*_s\bar D^*_s \,.
\nn
\eea
$D$-function for two-pole amplitude is equal to:
\bea
&&D_1\ =\ d_1+D_1\,L_{11}\,d_1+D_2\,L_{21}\,d_1\,,
\\ \nn
&&D_2\ =\ d_2+D_1\,L_{12}\,d_2+D_2\,L_{22}\,d_2\,,
\eea
so the multi-channel two-pole amplitude reads:
\bea \label{}
&&  A_{(X\to \alpha)}=
g_{(X\to 1)}D_{1}g_{(1\to\alpha)}+g_{(X\to 2)}D_{2}g_{(2\to\alpha)}=
\\
&&
g_{(X\to 1)}\frac{1}{\Delta}\bigg[d_{1}(1-g^2L_{22}d_2)
+d_2L_{21}d_1\bigg]g_{(1\to\alpha)}
+
g_{(X\to 2)}\frac{1}{\Delta}\bigg[d_{2}(1-L_{11}d_1)
+d_1L_{12}d_2\bigg]g_{(2\to\alpha)}\,,
\nn
\\
&&
L_{if}=\sum\limits_{\alpha'}
\int\limits_{(M_a+M_b)^2}^{+\infty}\frac{ds'}{\pi}
\frac{g_{i\alpha'} \rho_{\alpha'}(s')g_{f\alpha'} }{s'-s-i0}\,.
\nn
\eea

\section{ Spin wave functions of the \boldmath$(cs\cdot \bar c\bar s)$ systems
and their decomposition into meson-meson space}
\setcounter{equation}{0}

For writing meson-meson loop diagrams we need to know the spin wave functions in meson-meson
space - recombination of four-quark wave functions into two-meson ones is given below.
To write meson spin convolutions, it is sufficient to know wave functions with one fixed
component; we use $J_z=0$ wave functions.

\subsubsection*{\boldmath$A_{(cs)}\cdot A_{(\bar c\bar s)}$ systems with $J^{PC}=2^{++}$:}

\bea
&&
\sqrt6\,\psi^{(2,0)}_{\left(A_{(cs)}A_{\left(\bar c\bar s\right)}\right)}\ =\
\left(A^{(\Uparrow)}_{(cs)}\cdot A^{(\Downarrow)}_{(\bar c\bar s)}+2\,A^{(0)}_{(cs)}\cdot A^{(0)}_{(\bar c\bar s)}+
A^{(\Downarrow)}_{(cs)}\cdot A^{(\Uparrow)}_{(\bar c\bar s)}\right)
\\ \nn
&&=
(c^\uparrow\bar c^\downarrow)(s^\uparrow\bar s^\downarrow)-(c^\uparrow\bar s^\downarrow)(s^\uparrow\bar c^\downarrow)+
(c^\uparrow\bar c^\uparrow)(s^\downarrow\bar s^\downarrow)-(c^\uparrow\bar s^\downarrow)(s^\downarrow\bar c^\uparrow)
\\ \nn
&&\,\,+
(c^\uparrow\bar c^\downarrow)(s^\downarrow\bar s^\uparrow)-(c^\uparrow\bar s^\uparrow)(s^\downarrow\bar c^\downarrow)+
(c^\downarrow\bar c^\uparrow)(s^\uparrow\bar s^\downarrow)-(c^\downarrow\bar s^\downarrow)(s^\uparrow\bar c^\uparrow)
\\ \nn
&&
\,\,+
(c^\downarrow\bar c^\downarrow)(s^\uparrow\bar s^\uparrow)-(c^\downarrow\bar s^\uparrow)(s^\uparrow\bar c^\downarrow)+
(c^\downarrow\bar c^\uparrow)(s^\downarrow\bar s^\uparrow)-(c^\downarrow\bar s^\uparrow)(s^\downarrow\bar c^\uparrow)
\\ \nn
&&=\left(2\psi^{(0)}\phi^{(0)}+\psi^{(\Uparrow)}\phi^{(\Downarrow)}+\psi^{(\Downarrow)}\phi^{(\Uparrow)}
-2D^{*+(0)}_s\,D^{*-(0)}_s-D^{*+(\Uparrow)}_s\,D^{*-(\Downarrow)}_s-D^{*+(\Downarrow)}_s\,D^{*-(\Uparrow)}_s\right)
\,.
\eea

\subsubsection*{\boldmath$A_{(cs)}\cdot A_{(\bar c\bar s)}$ systems with $J^{PC}=1^{+-}$:}

\bea
&&\sqrt2\,\psi^{(1,0)}_{\left(A_{(cs)}A_{(\bar c\bar s)}\right)}\ =\
\left(A^{(\Uparrow)}_{(cs)}\cdot A^{(\Downarrow)}_{(\bar c\bar s)}
-A^{(\Downarrow)}_{(cs)}\cdot A^{(\Uparrow)}_{(\bar c\bar s)}\right)
\\ \nn
&&=
(c^\uparrow\bar c^\downarrow)(s^\uparrow\bar s^\downarrow)-(c^\uparrow\bar s^\downarrow)(s^\uparrow\bar c^\downarrow)-
(c^\downarrow\bar c^\uparrow)(s^\downarrow\bar s^\uparrow)+(c^\downarrow\bar s^\uparrow)(s^\downarrow\bar c^\uparrow)
\\ \nn
&&=\left(\psi^{(0)}\eta+\eta_c\phi^{(0)}
-D^{*+(0)}_s\,D^{-}_s-D^{+}_s\,D^{*-(0)}_s\right)
\,,
\eea

\subsubsection*{\boldmath$A_{(cs)}\cdot A_{(\bar c\bar s)}$ systems with $J^{PC}=0^{++}$:}

\bea
&&\sqrt3\,\psi^{(0,0)}_{\left(A_{(cs)}A_{(\bar c\bar s)}\right)}\ =\
\left(A^{(\Uparrow)}_{(cs)}\cdot A^{(\Downarrow)}_{(\bar c\bar s)}
-A^{(0)}_{(cs)}\cdot A^{(0)}_{(\bar c\bar s)}+
A^{(\Downarrow)}_{(cs)}\cdot A^{(\Uparrow)}_{(\bar c\bar s)}\right)\
\\ \nn
&&
=\quad
(c^\uparrow\bar c^\downarrow)(s^\uparrow\bar s^\downarrow)-(c^\uparrow\bar s^\downarrow)(s^\uparrow\bar c^\downarrow)+
(c^\downarrow\bar c^\uparrow)(s^\downarrow\bar s^\uparrow)-(c^\downarrow\bar s^\uparrow)(s^\downarrow\bar c^\uparrow)
\\ \nn
&&-\frac12\left[
(c^\uparrow\bar c^\uparrow)(s^\downarrow\bar s^\downarrow)-(c^\uparrow\bar s^\downarrow)(s^\downarrow\bar c^\uparrow)+
(c^\uparrow\bar c^\downarrow)(s^\downarrow\bar s^\uparrow)-(c^\uparrow\bar s^\uparrow)(s^\downarrow\bar c^\downarrow)
\right.
\\ \nn
&&\quad+\left.
(c^\downarrow\bar c^\uparrow)(s^\uparrow\bar s^\downarrow)-(c^\downarrow\bar s^\downarrow)(s^\uparrow\bar c^\uparrow)+
(c^\downarrow\bar c^\downarrow)(s^\uparrow\bar s^\uparrow)-(c^\downarrow\bar s^\uparrow)(s^\uparrow\bar c^\downarrow)
\right]
\\ \nn
&&=\frac12\left(\psi^{(0)}\phi^{(0)}+3\,\eta_c\eta-\psi^{(\Uparrow)}\phi^{(\Downarrow)}-\psi^{(\Downarrow)}\phi^{(\Uparrow)}
\right.
\\ \nn
&&\left.
-D^{*+(0)}_s\,D^{*-(0)}_s-3\,D^{+}_s\,D^{-}_s+D^{*+(\Uparrow)}_s\,D^{*-(\Downarrow)}_s+D^{*+(\Downarrow)}_s\,D^{*-(\Uparrow)}_s\right)
\,.
\eea

\subsubsection*{\boldmath$S_{(cs)}\cdot A_{(\bar c\bar s)}$ and
$A_{(cs)}\cdot S_{(\bar c\bar s)}$ systems with $J^{PC}=1^{++}$ and $J^{PC}=1^{+-}$:}

\bea
&&2\,\psi^{(1,0)}_{\left(S_{(cs)}A_{(\bar c\bar s)}\right)}\ =\
2\, \left(S_{(cs)}\cdot A^{(0)}_{(\bar c\bar s)}\right)
\\ \nn
&&=
 (c^\uparrow\bar c^\uparrow)(s^\downarrow\bar s^\downarrow)
-(c^\uparrow\bar s^\downarrow)(s^\downarrow\bar c^\uparrow)
+(c^\uparrow\bar c^\downarrow)(s^\downarrow\bar s^\uparrow)
-(c^\uparrow\bar s^\uparrow)(s^\downarrow\bar c^\downarrow)
\\ \nn
&&
\,\,
-(c^\downarrow\bar c^\uparrow)(s^\uparrow\bar s^\downarrow)
-(c^\downarrow\bar s^\downarrow)(s^\uparrow\bar c^\uparrow)
-(c^\downarrow\bar c^\downarrow)(s^\uparrow\bar s^\uparrow)
+(c^\downarrow\bar s^\uparrow)(s^\uparrow\bar c^\downarrow)
\\ \nn
&&=\left(-\psi^{(0)}\eta+\eta_c\phi^{(0)}+\psi^{(\Uparrow)}\phi^{(\Downarrow)}-\psi^{(\Downarrow)}\phi^{(\Uparrow)}
\right.
\\ \nn
&&\left.
+D^{*+(0)}_s\,D^{-}_s-D^{+}_s\,D^{*-(0)}_s-D^{*+(\Uparrow)}_s\,D^{*-(\Downarrow)}_s+D^{*+(\Downarrow)}_s\,D^{*-(\Uparrow)}_s\right)
\,,
\eea

\bea
&&
2\,\psi^{(1,0)}_{\left(A_{(cs)}S_{(\bar c\bar s)}\right)}\ =\
2\, \left(A^{(0)}_{(cs)}\cdot S_{(\bar c\bar s)}\right)
\\ \nn
&&=
 (c^\uparrow\bar c^\uparrow)(s^\downarrow\bar s^\downarrow)
-(c^\uparrow\bar s^\downarrow)(s^\downarrow\bar c^\uparrow)
-(c^\uparrow\bar c^\downarrow)(s^\downarrow\bar s^\uparrow)
+(c^\uparrow\bar s^\uparrow)(s^\downarrow\bar c^\downarrow)
\\ \nn
&&
\,\,
+(c^\downarrow\bar c^\uparrow)(s^\uparrow\bar s^\downarrow)
-(c^\downarrow\bar s^\downarrow)(s^\uparrow\bar c^\uparrow)
-(c^\downarrow\bar c^\downarrow)(s^\uparrow\bar s^\uparrow)
+(c^\downarrow\bar s^\uparrow)(s^\uparrow\bar c^\downarrow)
\\ \nn
&&=\left(\psi^{(0)}\eta-\eta_c\phi^{(0)}
+\psi^{(\Uparrow)}\phi^{(\Downarrow)}-\psi^{(\Downarrow)}\phi^{(\Uparrow)}
\right.
\\ \nn
&&\left.
+D^{*+(0)}_s\,D^{-}_s-D^{+}_s\,D^{*-(0)}_s+D^{*+(\Uparrow)}_s\,D^{*-(\Downarrow)}_s-D^{*+(\Downarrow)}_s\,D^{*-(\Uparrow)}_s\right)
\,.
\eea

\subsubsection*{The \boldmath$S_{(cs)}\cdot S_{(\bar c\bar s)}$ system with
$J^{PC}=0^{++}$:}

\bea
&& 2\,\psi^{(0,0)}_{\left(A_{(cs)}A_{(\bar c\bar s)}\right)}\ =\
2\,\left(S_{(cs)}\cdot S_{(\bar c\bar s)}\right)
\\ \nn
&&=
(c^\uparrow\bar c^\uparrow)(s^\downarrow\bar s^\downarrow)-(c^\uparrow\bar s^\downarrow)(s^\downarrow\bar c^\uparrow)-
(c^\uparrow\bar c^\downarrow)(s^\downarrow\bar s^\uparrow)+(c^\uparrow\bar s^\uparrow)(s^\downarrow\bar c^\downarrow)
\\ \nn
&&
\,\,-
(c^\downarrow\bar c^\uparrow)(s^\uparrow\bar s^\downarrow)+(c^\downarrow\bar s^\downarrow)(s^\uparrow\bar c^\uparrow)+
(c^\downarrow\bar c^\downarrow)(s^\uparrow\bar s^\uparrow)-(c^\downarrow\bar s^\uparrow)(s^\uparrow\bar c^\downarrow)
\\ \nn
&&=\left(-\psi^{(0)}\phi^{(0)}+\eta_c\eta+\psi^{(\Uparrow)}\phi^{(\Downarrow)}+\psi^{(\Downarrow)}\phi^{(\Uparrow)}
\right.
\\ \nn
&&\left.
-D^{*+(0)}_s\,D^{*-(0)}_s+\,D^{+}_s\,D^{-}_s+D^{*+(\Uparrow)}_s\,D^{*-(\Downarrow)}_s+D^{*+(\Downarrow)}_s\,D^{*-(\Uparrow)}_s\right)
\,.
\eea

\end{document}